\documentclass[12pt]{article}
 \usepackage{epsfig}
\columnwidth\textwidth

\begin{document}

\newcommand{\be}{\begin{equation}}
\newcommand{\ee}{\end{equation}}
\newcommand{\beann}{\begin{eqnarray*}}
\newcommand{\eeann}{\end{eqnarray*}}
\newcommand{\bea}{\begin{eqnarray}}
\newcommand{\eea}{\end{eqnarray}}
\newcommand{\nn}{\nonumber}
\newcommand{\ben}{\begin{enumerate}}
\newcommand{\een}{\end{enumerate}}
\newcommand{\lb}{\label}
\newtheorem{df}{Definition}
\newtheorem{thm}{Theorem}
\newtheorem{lem}{Lemma}
\newtheorem{prop}{Proposition}
\begin{titlepage}

\noindent
\vspace*{1cm}
\begin{center}
{\large\bf SINGULARITY AVOIDANCE BY COLLAPSING SHELLS} \\[0.5cm]
{\large\bf IN QUANTUM GRAVITY}\footnote{This essay
 received an "honorable mention" in the 2001 Essay
    Competition of the Gravity Research Foundation -- Ed.}  

\vskip 1cm
{\bf Petr H\'{a}j\'{\i}\v{c}ek} 
\vskip 0.4cm
Institut f\"ur Theoretische Physik, \\
Universit\"at Bern, \\
Sidlerstrasse 5, 3012 Bern, Switzerland.\\
\vskip 0.7cm

{\bf Claus Kiefer} 
\vskip 0.4cm
Institut f\"ur Theoretische Physik,\\ Universit\"{a}t zu K\"oln, \\
Z\"ulpicher Str.~77,
50937 K\"oln, Germany.\\
\vspace{1cm}

\nopagebreak[4]

\begin{abstract}
We discuss a model describing exactly
 a thin spherically symmetric
shell of matter with zero rest mass. We derive the reduced
formulation of this system in which the variables are
embeddings, their conjugate momenta, and Dirac observables.
A non-perturbative quantum theory of this model is then
constructed, leading to a unitary dynamics. As a consequence of
unitarity, the classical singularity is fully avoided in the
quantum theory.
\end{abstract}
\vskip 0.4cm

\end{center}

\end{titlepage}

The construction of a full, non-perturbative, quantum theory
of gravity is one of the main open problems in
physics. To achieve this goal, one is basically
confronted with two options. On the one hand, one can attempt
to construct first a full quantum theory of gravity
and then derive interesting physical consequences.
Superstring theory and the Ashtekar
approach to canonical quantum gravity constitute two examples.
On the other hand, one can try to construct first
models which can be exactly quantized and which can therefore
serve as a guide towards the full, unknown, theory.
Such models can even have a direct bearing on concrete
physical situations. This would be reminiscent of the first
calculations of atomic spectra which were feasible
without the knowledge of full quantum elctrodynamics.

In our essay we shall follow the second route.
The model that we shall discuss is a spherically-symmetric
thin shell consisting of particles with zero rest mass
(``lightlike shell'').
According to the classical theory of general relativity,
such a shell can collapse to form a black hole.
Alternatively, it can emerge from a white hole and expand to
infinity. The latter possibility is usually excluded
for thermodynamical reasons. 

The classical theory predicts that a genuine gravitational
collapse leads to spacetime singularities \cite{H-E}.
A special feature is the occurrence of a horizon during
the collapse, a region from within no information can
escape to the outside. It is a general expectation that
a quantum theory of gravity can cure this situation,
i.e., can lead to a singularity-free geometry. In fact, we shall
show in this essay that a quantum theory for the 
lightlike shell leading to a singularity-free situation
 can be rigorously constructed.
This will be a consequence of
the unitary dynamics. 

For the discussion of the model, we shall employ
the approach of reduced quantization. In this approach,
the variables can be neatly separated into pure gauge
degrees of freedom (so-called embedding variables),
their canonical momenta, and physical degrees of freedom
\cite{canada}. The general existence
of this ``Kucha\v{r} decomposition'' was shown in \cite{H-K}
by making a transformation to the usual (ADM) phase space
of general relativity.
In the construction, the notion of a background
manifold plays a crucial role because one must define
points by the choice of coordinates on some fixed
manifold (a priori, spacetime points have no intrinsic
meaning because they can be moved around by the
diffeomorphism group). 

The example of the lightlike shell constitutes the first
application of this method \cite{HK}. For the coordinates on
the fixed manifold, double-null coordinates $U$ and $V$
are chosen on the background manifold
${\mathcal M}={\mathbf R}_+ \times {\mathbf R}$
(being effectively two-dimensional due to spherical symmetry).
In these coordinates (which will play the role of
the embedding variables),
 the metric has the form
\begin{equation}
  ds^2  =  -A(U,V)dUdV + R^2(U,V)(d\vartheta^2 +
  \sin^2\vartheta d\varphi^2)\ .
\label{KRV}
\end{equation}
{}From the demand that the metric be regular at the center
and continuous at the shell, the coefficients $A$ and $R$
are uniquely defined for any physical situation defined
by the variables $M$ (the energy of the shell), $\eta$
(being +1 for an outgoing shell and -1 for an ingoing shell),
and $w$ (the location of the shell, where $w=u$ for the outgoing
and $w=v$ for the ingoing case). 
The Penrose diagram for the outgoing shell is shown in 
Figure~1.
It is important to note that the background manifold
possesses a unique asymptotic region with
${\mathcal J}^-$ defined by $U\to-\infty$ and
${\mathcal J}^+$ by $V\to+\infty$.
\begin{figure}[htb]
\begin{center}
\epsfig{file=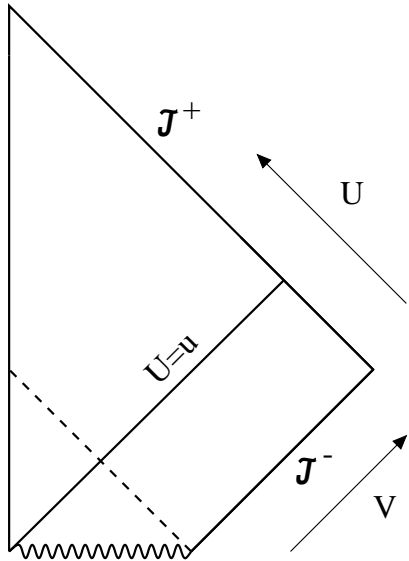}
\end{center}
\caption{Penrose diagram for the outgoing shell in the classical
         theory.
         The shell is at $U=u$.}
\end{figure} 

We shall first transform the classical
theory of the shell into the formulation corresponding to
the Kucha\v{r} decomposition and then construct the
quantum theory. The standard (ADM) formulation of the shell
was studied in \cite{L-W-F}.
One can perform an explicit transformation of these
ADM variables into the new variables
 $u$ and $v$, their momenta $p_u$ and $p_v$,
the embedding variables
$U(\rho)$, $V(\rho)$, and their momenta
\cite{HK}. 
The result is the action
\begin{equation}
  S = \int d\tau\left(p_u\dot{u} + p_v\dot{v} - np_up_v\right)
  + \int d\tau\int_0^\infty d\rho(P_U\dot{U} + P_V\dot{V} - H)\ ,
\label{KD}
\end{equation}
where $H = N^UP_U + N^VP_V$, and $n$, $N^U(\rho)$, and $N^V(\rho)$ are
Lagrange multipliers. The first term in (\ref{KD}) contains the physical
variables, while the second term contains the gauge variables.  Observe that
the Poisson algebra of the chosen set of observables $p_u$ and $u$ for $\eta =
+1$ as well as $p_v$ and $v$ for $\eta = -1$ is gauge invariant in spite of
the fact that it has been obtained by a calculation based on a gauge choice
(the double-null coordinates $U$ and $V$). This implies that our construction
of the quantum mechanics will also be gauge invariant.  A crucial point is
that the new phase space has non-trivial boundaries:
\begin{equation}
  p_u \leq 0,\quad p_v \leq 0\ ,
\quad
 \frac{-u+v}{2} > 0\ .
\label{78}
\end{equation}
The boundary defined by the last inequality
is due to the classical singularity. 

The system has now been brought into a form where it can
be subject to quantization \cite{PH}. The restrictions 
(\ref{78}) suggest the use of the
so-called group-quantization method \cite{Isham}.
This method leads automatically to self-adjoint
operators for the observables. A complete system of Dirac observables
is given by $p_u$, $p_v$, as well as $up_u$ and
$vp_v$. They thus commute with the constraint
$p_up_v$.
 The Hilbert space is constructed from
complex functions $\psi_u(p)$ and $\psi_v(p)$, where
$p\in [0,\infty)$. The scalar product is defined by
\be 
  (\psi_u,\phi_u) := \int_0^\infty\ \frac{dp}{p}\
    \psi^*_u(p)\phi_u(p) \label{Hilbert}
\ee
(similarly for $\psi_v(p)$). To handle the inequalities
(\ref{78}) it is useful to perform the following
canonical transformation:
\begin{eqnarray}
  t & = & (u+v)/2, \qquad r = (-u+v)/2,
\label{tr} \\
  p_t & = & p_u + p_v, \qquad \;\; p_r = -p_u + p_v\ .
\label{ptpr}
\end{eqnarray}
Upon quantization, one obtains the operator $-\hat{p}_t$
which is self-adjoint and has a positive spectrum. It is the
generator of time evolution and corresponds to the energy operator
$\hat{M}$.
 Since $r$ is not a Dirac observable,
it cannot directly be transformed into a quantum observable. 
It turns out that the following construction is 
useful \cite{PH}:
\begin{equation}
 \hat{r}^2 := 
  -\sqrt{p}\frac{d^2}{dp^2}\frac{1}{\sqrt{p}}\ .
\label{rsym}
\end{equation}
This is essentially a Laplacian and corresponds to a
concrete choice of factor ordering. It is a symmetric operator
which can be extended to a self-adjoint operator.
In this process, one is naturally led to the
following eigenfunctions of $\hat{r}^2$:
\begin{equation}
  \psi(r,p) := \sqrt{\frac{2p}{\pi}}\sin rp\ ,\quad r \geq 0 \ .
\label{rsa}
\end{equation}
One can also construct an operator $\hat{\eta}$ that classically
would correspond to the direction of motion of the shell.

Now the basic formalism of the quantum theory is set and one
can start to study concrete physical applications.
We want to describe the dynamics of the
shell by the evolution of a narrow wave packet. 
We take for $t=0$ the following family
of wave packets:
\be
  \psi_{\kappa\lambda}(p) :=
   \frac{(2\lambda)^{\kappa+1/2}}{\sqrt{(2\kappa)!}}
  p^{\kappa+1/2}e^{-\lambda p}\ ,
\ee
where $\kappa$ is a positive integer,
and $\lambda$ is a positive number with
dimension of length. 
By an appropriate choice of these constants one can
prescribe the expectation value of the energy and its
variation. A sufficiently narrow wave packet can thus be
constructed.

Since the time evolution of the packet is generated by $-\hat{p}_t$,
one finds
\be
  \psi_{\kappa\lambda}(t,p) = \psi_{\kappa\lambda}(p) e^{-ipt}\ .
\ee
More interesting is the evolution of the wave packet
in the $r$-representation. This is obtained by the
integral transform (\ref{Hilbert})
 of $\psi_{\kappa\lambda}(t,p)$
with respect to the eigenfunctions (\ref{rsa}).
It leads to the exact result
\begin{equation}
  \Psi_{\kappa\lambda}(t,r) = \frac{1}{\sqrt{2\pi}}
  \frac{\kappa!(2\lambda)^{\kappa+1/2}}{\sqrt{(2\kappa)!}}
  \left[\frac{i}{(\lambda +it +ir)^{\kappa+1}}
   - \frac{i}{(\lambda +it-ir)^{\kappa+1}}\right]. 
\label{17:1}
\end{equation}
One interesting consequence can be immediately drawn:
\be
  \lim_{r\rightarrow 0}\Psi_{\kappa\lambda}(t,r)= 0\ .
\ee 
This means that the probability to find the shell at
vanishing radius is zero! In this sense the singularity is avoided
in the quantum theory. We emphasize that this is not
a consequence of a certain boundary condition -- it is a
consequence of the {\em unitary evolution}. If the wave function
vanishes at $r=0$ for $t\to-\infty$ (asymptotic condition
of ingoing shell), it will continue to vanish at $r=0$
for all times. It follows from (\ref{17:1}) that the quantum shell 
bounces and re-expands. Hence, no absolute event horizon can form,
in contrast to the classical theory. However, an object that is locally
similar to a black hole is not excluded by our results. In this way, the
observational support for black holes is not contradicted.

Most interestingly, an essential part of the wave packet can
even be squeezed below the expectation value of its
Schwarzschild radius. This is achieved if the 
expectation value of the energy fulfills the condition
\begin{equation}
  \langle\hat{M}\rangle > \frac{\lambda M_P}{\sqrt{2\pi}}M_P\ ,
\label{squee2}
\end{equation}
where $M_P$ denotes the Planck mass,
and $\lambda M_P\gg 1$ holds \cite{PH}.
 The wave packet can thus be squeezed
below its Schwarzschild radius if its energy is much bigger
than the Planck energy -- a genuine quantum effect.

How can this behavior be understood? 
The unitary dynamics ensures that the ingoing quantum shell
develops into a {\em superposition} of ingoing and outgoing
shell if the region is reached where in the classical theory
a singularity would form. In other words, the singularity is
avoided by destructive interference in the quantum theory.
This is similar to the quantum-cosmological example
of \cite{KZ} where a superposition of a black hole with
a white hole leads to a singularity-free quantum universe.
Also here, the horizon becomes a superposition
of ``black hole'' and ``white hole'' -- its ``grey'' nature 
can be characterized by the expectation value of the
operator $\hat{\eta}$ (a black-hole horizon would correspond
to the value -1 and a white-hole horizon to the value +1).
We emphasize that in this scenario no information-loss
paradox would ever arise if such a behavior occurred for all
collapsing matter (which sounds reasonable).
 In the same way, the principle of
cosmic censorship would be implemented, since no naked
singularities (in fact, no singularities at all) would 
form.

To summarize, the non-perturbative study of the
above example demonstrates how the process of gravitational
collapse may be viewed in quantum gravity. Whether the full,
elusive, theory will be in accordance with this picture
is of course an open question.

\end{document}